\documentclass[english,aps,preprint]{revtex4}
\usepackage[T1]{fontenc}
\usepackage[latin9]{inputenc}
\usepackage{color}
\usepackage{babel}
\usepackage{textcomp}
\usepackage{amsmath}
\usepackage{amssymb}
\usepackage{graphicx}
\usepackage[unicode=true]
 {hyperref}

\makeatletter

\newcommand*\LyXThinSpace{\,\hspace{0pt}}

\@ifundefined{textcolor}{}
{%
 \definecolor{BLACK}{gray}{0}
 \definecolor{WHITE}{gray}{1}
 \definecolor{RED}{rgb}{1,0,0}
 \definecolor{GREEN}{rgb}{0,1,0}
 \definecolor{BLUE}{rgb}{0,0,1}
 \definecolor{CYAN}{cmyk}{1,0,0,0}
 \definecolor{MAGENTA}{cmyk}{0,1,0,0}
 \definecolor{YELLOW}{cmyk}{0,0,1,0}
}

\makeatother

\begin{document}
\begin{center}
\textbf{\large{}Thermal divergences of quantum measurement engine}{\large\par}
\par\end{center}
\author{Shanhe Su}
\thanks{The two authors contributed equally to this work.}
\author{$^{\dagger}$Zhiyuan Lin}
\thanks{The two authors contributed equally to this work.}
\author{Jincan Chen}
\thanks{Electronic addresses: sushanhe@xmu.edu.cn;jcchen@xmu.edu.cn}
\affiliation{Department of Physics, Xiamen University, Xiamen 361005, People's
Republic of China}
\begin{abstract}
A quantum engine fueled by quantum measurement is proposed. Under
the finite-time adiabatic driving regime, the conversion of heat to
work is realized without the compression and expansion of the resonance
frequency. \textcolor{black}{The work output, quantum heat, and efficiency
are derived, highlighting the important role of the thermal divergence
recently reappearing in open quantum systems. The key problem of how
the measurement basis can be optimized to enhance the performance
is solved by connecting the thermal divergence to the nonequilibrium
free energy and entropy. }The spin-engine architecture offers a comprehensive
platform for future investigations of extracting work from quantum
measurement. 
\end{abstract}
\maketitle

\subsection*{1. Introduction}

The developments of nuclear magnetic resonance, trapped ions, superconducting
circuits, and quantum state tomography facilitate access to the measurement
and manipulation of spin ensembles \cite{key-1,key-2}. These techniques
have been employed to verify how the laws of thermodynamic emerge
in the quantum domain \cite{key-3,key-4}. The existence of the arrow
of time in a thermodynamically irreversible process has been demonstrated
by observing a fast quenched dynamics of an isolated spin-1/2 system
\cite{key-5}. The reversal of heat flow from a cold spin to a hot
spin is possible due to the initial quantum correlation between the
spins \cite{key-6}. An experimental reconstruction of the dynamics
of a closed quantum spin showed that the statistical distribution
of work satisfies the Tasaki-Crooks theorem and the Jarzynski equality
\cite{key-7}. By integrating the two-point measurement, unitary evolution,
and feedback in the control of a spin-1/2 of the $^{13}C$ nucleus,
the fluctuation relation and the non-equilibrium entropy production
in the presence of information have been observed \cite{key-8}. 

More interestingly, the implementations of quantum machines based
on spin systems unlock the potential role of quantum effects in the
processes of energy conversion and transport \cite{key-9,key-10,key-11,key-12}.
Peterson et al. experimentally implemented a spin engine with an efficiency
for work extraction close to Carnot's efficiency \cite{key-13}. Assis
et al. densigned a quantum engine with efficiency higher than the
Otto limit, since a reservoir is prepared at a negative effective
temperature by inverting the population of a huge nuclear hydrogen
spin system \cite{key-14}. Klatzow and co-workers proved that a coherent
superposition enables an engine to produce more power than the equivalent
classical counterpart by using an ensemble of nitrogen vacancy centers
in diamond as the working substance \cite{key-15}. Considering the
non-Markovian effect, Shirai et al. identified a new definition of
work including the energy cost of detaching the spin from the reservoir
\cite{key-16,key-17}. 

A natural question that arises is to unveil the role of quantum measument
in the energy conversion. When quantum measurement collides with spin
ensembles, will it be feasible to create the methodology of work extraction
in a more flexible and efficient way? It was recently reported that
quantum measurement changes the average energy of a quantum system,
when the measured observable and the system Hamiltonian do not commute
\cite{key-18,key-19,key-20}. Instead of the stochastic thermal fluctuation
from a hot bath, work can be extracted from the stochastic quantum
fluctuation induced by the measurement process in quantum Maxwell\textquoteright s
demon engine \cite{key-21}. For a two-stroke two-spin device, invasive
quantum measurement was regarded as a resource to power the refrigeration
\cite{key-22}. The idea of using the energy provided by the measurement
apparatus as the fuel is a fundamental aspect of quantum engines.
However, the advantages of adiabatic processes experiencing time-dependent
Hamiltonians in quantum measurement engines have not been well revealed.
Meanwhile, no simple expressions associated with the heat and work
have been obtained so far. The efforts to address these issues will
provides general information on how the measurement basis affects
the performance of an engine. \textcolor{black}{Thus, the physical
implication of the thermal divergence needs to be further excavated
because of its simplicity and importance in relating the nonequilibrium
free energy and entropy of a quantum state. }

In this work, a four-stroke engine, where the quantum measument and
thermalization processes are connected by two thermodynamic adiabatic
processes, is built. The quantum measument process will provide the
input energy to ignite the engine. A time-dependent evolution of a
spin driven by a rotating magnetic field is applied in the adiabatic
processes. Analytical expressions that relate the thermodynamic quantities
regarding the thermal divergence, nonequilibrium free energy, and
entropy to the heat and work along the cycle will be presented. How
the angles of the measurement basis on the bloch sphere influence
the performance will be revealed through the direct relations among
the work, efficiency, and the newly defined thermodynamic quantities. 

\subsection*{2. The quantum measurement engine}

The schematic of the quantum-measurement engine employing a single
spin as the working substance is illustrated in Fig. 1. The top and
bottom of the loop represent a pair of quasi-parallel and isentropic
processes. An isentropic process implies that the von Neumann entropy
of the quantum system remains constant in time because of its invariance
through unitary evolution, and the system is impermeable to heat during
this process. The left and right sides of the loop are a pair of parallel
isochoric processes with constant external fields. Heat flows into
the cycle through the right quantum measurement process and a part
of heat is dumped into the heat sink through the left thermalization
process. 

\noindent 
\begin{figure}
\noindent \begin{centering}
\includegraphics[scale=0.35]{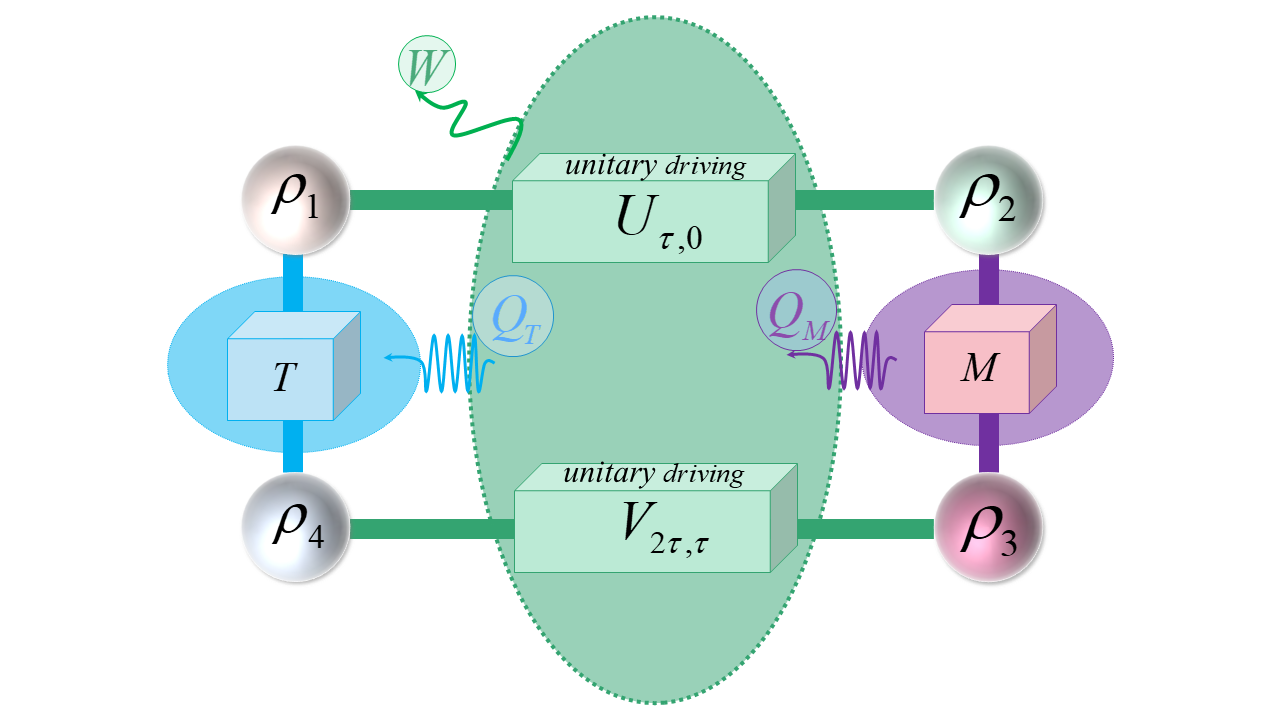}
\par\end{centering}
\caption{Schematic of the quantum-measurement engine consisting of two adiabatic
processes, a quantum measuement process, and a thermalization process.}
\end{figure}

The system is in a pseudo-thermal state $\rho_{1}=e^{-\beta H_{1}}/\mathrm{Tr}\left(e^{-\beta H_{1}}\right)$
at inverse temperature $\beta$ and Hamiltonian $H_{1}=\frac{\hbar\omega}{2}\sigma_{z}$
, where $\hbar$ is the reduced Planck constant, $\omega$ denotes
the resonance frequency, and $\sigma_{i}\left(i=x,y,z\right)$ are
the Pauli spin operators. During the first stroke (from $t=0$ to
$t=\tau$), the system undergoes a thermally isolated transformation.
A time-modulated radiofrequency field resonant with the spin makes
Hamiltonian depend on time, which is described by the effective Hamiltonian
$H_{I}(t)=\frac{\hbar\omega}{2}\left(\cos\frac{\pi t}{2\tau}\sigma_{z}+\textrm{sin}\frac{\pi t}{2\tau}\sigma_{x}\right)$.
At the end of the stroke, the system attains the state $\rho_{2}=U_{\tau,0}\rho_{1}U_{\tau,0}^{\dagger}$
with $U_{\tau,0}=\mathcal{T}e^{-\frac{i}{\hbar}\int_{0}^{\tau}H_{I}\left(t\right)dt}$
denoting the time-evolution operator and $\mathcal{T}$ being the
time-ordering operator, while the Hamiltonian at time $t=\tau$ becomes
$H_{2}=\frac{\hbar\omega}{2}\sigma_{x}$. The system consumes an amount
of work $W_{1}=\textrm{Tr}\left(\rho_{2}H_{2}-\rho_{1}H_{1}\right)$.

In the second stroke, an instantaneous projective measurement is the
testing on the system. For a positive operator-valued measure (POVM)
with a set of orthogonal projectors ${\displaystyle \left\{ \left|\chi_{1}\right\rangle \left\langle \chi_{1}\right|,\left|\chi_{2}\right\rangle \left\langle \chi_{2}\right|\right\} }$,
the post-measurement state $\rho_{2}$ is updated to $\rho_{3}=\sum_{k}\pi_{k}\rho_{2}\pi_{k}$.
In this study, we can perform a measurement that projects the spin
onto the basis $\left|\chi_{1}\right\rangle =e^{-i\varphi}\sin\frac{\alpha}{2}\left|\uparrow\right\rangle -\cos\frac{\alpha}{2}\left|\downarrow\right\rangle $
and $\left|\chi_{2}\right\rangle =\cos\frac{\alpha}{2}\left|\uparrow\right\rangle +e^{i\varphi}\sin\frac{\alpha}{2}\left|\downarrow\right\rangle $
. The measurement basis corresponds to measuring the state of the
spin in a particular direction. The parameters $\alpha$ and $\varphi$
are, respectively, interpreted as the colatitude with respect to the
$z$-axis and the longitude with respect to the $x$-axis in the Bloch
sphere representation, where ${\displaystyle 0\leq\alpha\leq\pi}$
and ${\displaystyle 0\leq\phi<2\pi}$. As the Hamiltonian $H_{3}$
of the system corresponding to state $\rho_{3}$ remains the same
as $H_{2}$, the measurement gives rise to the energy change of the
spin $Q_{M}=\textrm{Tr}\left[H_{2}\left(\rho_{3}-\rho_{2}\right)\right]$,
which will be regarded as the fuel energy of the engine.

During the third stroke (from $t=\tau$ to $t=2\tau$), the system
evolutes unitarily according to the Hamiltonian $H_{II}(t)=\frac{\hbar\omega}{2}\left[\cos\frac{\pi(2\tau-t)}{2\tau}\sigma_{z}+\textrm{sin}\frac{\pi(2\tau-t)}{2\tau}\sigma_{x}\right]$,
which is a time-reversed protocol of the first adiabatic stroke. The
system Hamiltonian is driven from $H_{3}$ to $H_{4}$, where $H_{4}$
is equivalent to the origin form $H_{1}$ of the cycle. The final
state of the second adiabatic stroke $\rho_{4}=V_{2\tau,\tau}\rho_{3}V_{2\tau,\tau}^{\dagger}$
with the unitary operator $V_{2\tau,\tau}=\mathcal{T}e^{-\frac{i}{\hbar}\int_{\tau}^{2\tau}H_{II}\left(t\right)dt}$.
The work performed by the exteral field is given by $W_{2}=\textrm{Tr}\left(\rho_{4}H_{4}-\rho_{3}H_{3}\right)$.
In the last stroke, the system with a time-independent Hamiltonian
is coupled to the bath at inverse temperature $\beta$ . After a sufficiently
long time, the system is restored to the initial Gibbs state $\rho_{1}$
and releases the heat $Q_{T}=\textrm{Tr}\left[H_{1}\left(\rho_{1}-\rho_{4}\right)\right]$
to the bath.

\subsection*{3. The roles of the thermal divergence, nonequilibrium free energy,
and entropy in the performance}

Now, we examine simple expressions for heat and work in thermodynamic
processes via establishing the concept of thermal divergence. The
quantum Kullback-Leibler-Umegaki divergence (also called relative
entropy) \cite{key-23,key-24} is a measure of the entropy of the
system at state $\rho$ relative to state $\sigma$, given by
\begin{equation}
D(\rho\|\sigma)=\operatorname{Tr}[\rho\left(\ln\rho-\ln\sigma\right)].\label{eq:KLU}
\end{equation}
When the reference state $\sigma$ is a thermal equilibrium state,
$D(\rho\|\sigma)$ is called the thermal divergence \cite{key-25}.
Thermodynamic quantities, such as heat and work, can be written as
functions of thermal divergences of quantum states. 

The thermal equilibrium state of a quantum system with Hamiltonians
$H_{i}$ in contact with a heat bath at inverse temperature $\beta$
is given by the Gibbs state $\sigma_{i}^{eq}=e^{-\beta H_{i}}/Z_{i}$
with the partition function $\mathrm{\mathit{Z_{i}}=Tr}\left(e^{-\beta H_{i}}\right)$.
Since $\ln\sigma_{i}^{eq}=-\beta\left(H_{i}-F_{i}^{\mathrm{eq}}\right)$
with $F_{i}^{\mathrm{eq}}=-\beta{}^{-1}\ln Z_{i}$ being the Helmholtz
free energy, the thermal divergence is then be rewritten as
\begin{equation}
D\left(\rho_{i}\|\sigma_{i}^{eq}\right)=\beta\left(\mathcal{E}_{i}-F_{i}^{\mathrm{eq}}\right)-S_{i}\text{,}\label{eq:div-energy-entropy}
\end{equation}
where $\mathcal{E}_{i}=\textrm{Tr}\left[H_{i}\rho_{i}\right]$ is
the internal energy depending on the expectation value of the system
Hamiltonian and $S_{i}=-\textrm{Tr}\left[\rho_{i}\ln\rho_{i}\right]$
is the von Neumann entropy of state $i$. The thermal divergence has
been regarded as a fundamental quantity in quantum thermodynamics
because of its connection with the variables of energy and entropy
\cite{key-8,key-25}. By introducing the nonequilibrium free energy
\cite{key-26} of a quantum system as
\begin{equation}
F_{i}=\mathcal{E}_{i}-S_{i}/\beta\text{,}
\end{equation}
the thermal divergence would measure the discrepancy between the nonequilibrium
and equilibrium free energies
\begin{equation}
D\left(\rho_{i}\|\sigma_{i}^{eq}\right)=\beta\left(F_{i}-F_{i}^{\mathrm{eq}}\right)\text{.}\label{eq:free energy difference}
\end{equation}
\textcolor{black}{In other words, the nonequilibrium free energy of
a state is higher than that of the corresponding equilibrium state
by an amount equal to the temperature times the thermal divergence.}

By using Eq. (\ref{eq:div-energy-entropy}), the net work (Appendix)
done on the system in a complete cycle is simplified as

\begin{align}
W & =\frac{1}{\beta}\left[D\left(\rho_{4}||\sigma_{4}^{eq}\right)+D\left(\rho_{2}||\sigma_{2}^{eq}\right)-D\left(\rho_{3}||\sigma_{3}^{eq}\right)\right]\nonumber \\
 & =F_{2}-F_{1}+F_{4}-F_{3}.\label{eq:W}
\end{align}
It is surprised to find that the net work depends only on thermal
divergences of quantum states. With the help of Eq. (\ref{eq:free energy difference}),
$W$ is simplified into the sum of the changes of the nonequilibrium
free energies in the two adiabatic processes. For the second equality,
one has considered $S_{1}=S_{2}$ and $S_{3}=S_{4}$ because of the
invariant of von Neumann entropy under a unitary evolution, and $F_{3}^{eq}=F_{2}^{eq}$
and $F_{4}^{eq}=F_{1}^{eq}=F_{1}$. For extracting energy from the
cycle, the net work $W$ are restricted to $W<0$. 

The fuel energy (Appendix) provided by the measurement process is
represented by

\begin{align}
Q_{M} & =\frac{1}{\beta}\left[D\left(\rho_{3}||\sigma_{3}^{eq}\right)-D\left(\rho_{2}||\sigma_{2}^{eq}\right)+\Delta S_{M}\right]\nonumber \\
 & =F_{3}-F_{2}+\frac{1}{\beta}\Delta S_{M},\label{eq:QM}
\end{align}
where $\Delta S_{M}=S_{3}-S_{2}$ is the entropy difference of states
$\rho_{3}$ and $\rho_{2}$ and the relation $F_{3}^{eq}=F_{2}^{eq}$
has been applied. The fuel energy relies on $D\left(\rho_{3}||\sigma_{3}^{eq}\right)-D\left(\rho_{2}||\sigma_{2}^{eq}\right)$
and $\Delta S_{M}$. \textcolor{black}{Because the thermalization
process interacting with a hot reservoir in a standard Otto cycle
has been replaced by a measurement protocol, the input energy $Q_{M}$
is called as quantum heat. The heat (Appendix)} absorded from the
bath in the thermalization process reads

\begin{align}
Q_{T} & =\frac{1}{\beta}\left[-D\left(\rho_{4}||\sigma_{4}^{eq}\right)+\Delta S_{T}\right]\nonumber \\
 & =F_{1}-F_{4}+\frac{1}{\beta}\Delta S_{T}\text{,}
\end{align}
where $\Delta S_{T}=S_{1}-S_{4}$ and the thermal divergence $D\left(\rho_{1}\|\sigma_{1}^{eq}\right)=0$
have been used. Since the initial and final states are identical in
a cycle, the system experiences no change in energy. As a result,
$W+Q_{M}+Q_{T}=0$, satisfying the first law of thermodynamics. The
thermal efficiency is the net work output divided by the quantum heat
provided by quantum measurement

\begin{align}
\eta & =\frac{-W}{Q_{M}}\nonumber \\
 & =\frac{D\left(\rho_{3}||\sigma_{3}^{eq}\right)-D\left(\rho_{4}||\sigma_{4}^{eq}\right)-D\left(\rho_{2}||\sigma_{2}^{eq}\right)}{D\left(\rho_{3}||\sigma_{3}^{eq}\right)-D\left(\rho_{2}||\sigma_{2}^{eq}\right)+\Delta S_{M}}\nonumber \\
 & =1+\frac{F_{1}-F_{4}-\frac{1}{\beta}\Delta S_{M}}{F_{3}-F_{2}+\frac{1}{\beta}\Delta S_{M}},
\end{align}
where $\Delta S_{T}=-\Delta S_{M}$.

\subsection*{4. T\textcolor{black}{he performance of the thermodynamic cycle as
an engine} }

Figs. 2(a) and 2(b) show the contour plots of the work output \textminus $W$
and efficiency $\eta$ versus the colatitude $\alpha$ and longitude
$\varphi$, parameterized in $\hbar\omega=0.5\textrm{peV}$, $\tau=8.4\textrm{\ensuremath{\mu}s}$,
and $\beta=1/\hbar\omega_{0}$. Quantum measurements act as a powerful
tool for manipulating quantum states and providing the fuel energy
to drive the engine. Interestingly, the performance of the cycle can
be enhanced by optimizing the direction of the measurement basis.
The contour plots exhibit two qualitatively different regimes separated
by $\varphi=\pi$. For $0\leq\varphi\leq\pi$, \textminus $W$ shows
maxima at $\alpha_{W}=1.10$ and $\varphi_{W}=1.77$, while $\eta$
achieves the peak at $\alpha_{\eta}=1.15$ and $\varphi_{\eta}=2.04$.
\textcolor{black}{In the range of} $\pi\leq\varphi\leq2\pi$, the
distributions of \textminus $W$ and $\eta$ as functions of $\alpha$
and $\varphi$ satisfy antisymmetry with respective to the axis $\alpha=\pi/2$
and translational invariance, i.e., $-W\left(\alpha,\varphi\right)=-W\left(\pi-\alpha,\varphi+\pi\right)$
and $\eta\left(\alpha,\varphi\right)=\eta\left(\pi-\alpha,\varphi+\pi\right)$. 

\noindent 
\begin{figure}
\noindent \begin{centering}
\includegraphics[scale=0.35]{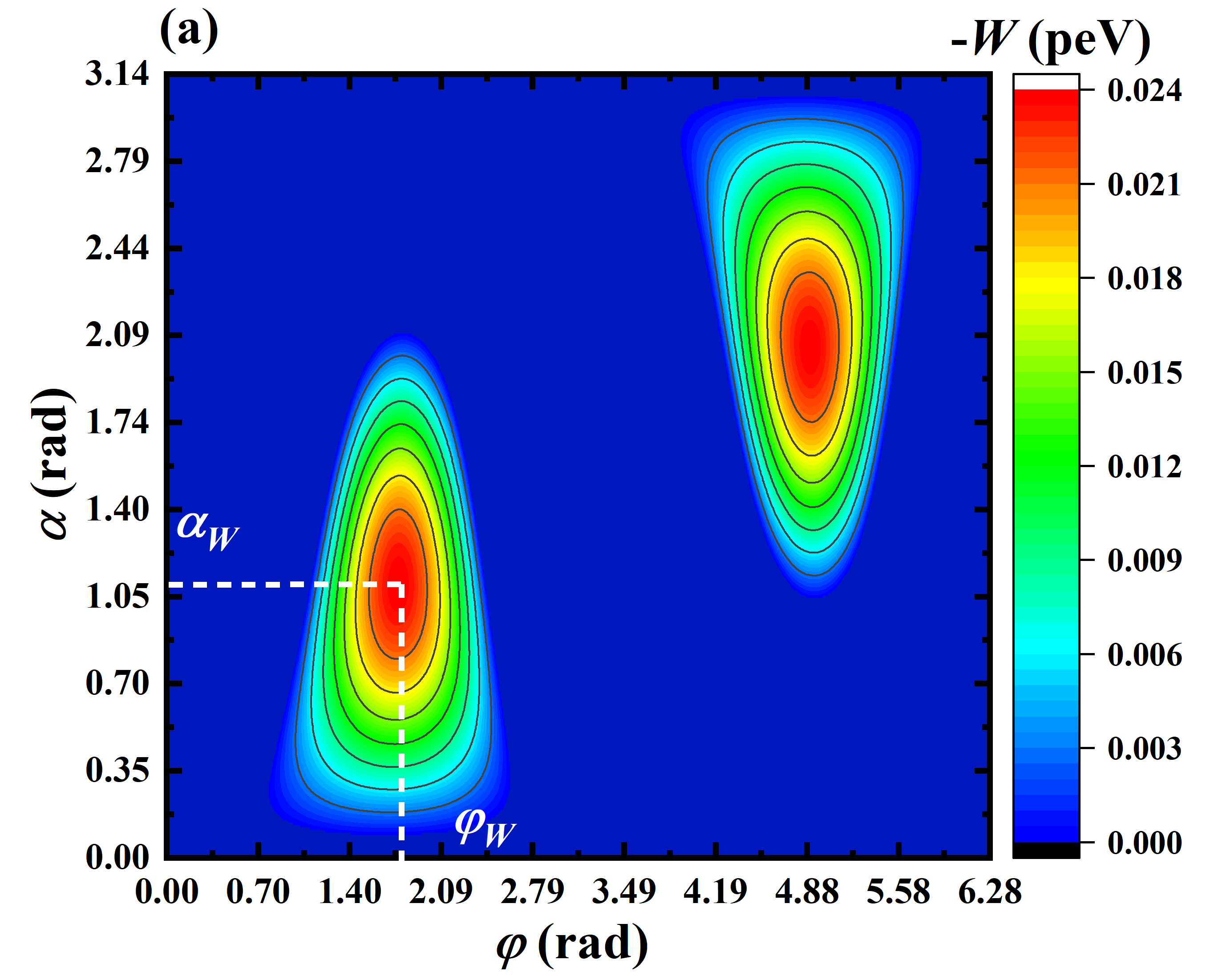}\includegraphics[scale=0.35]{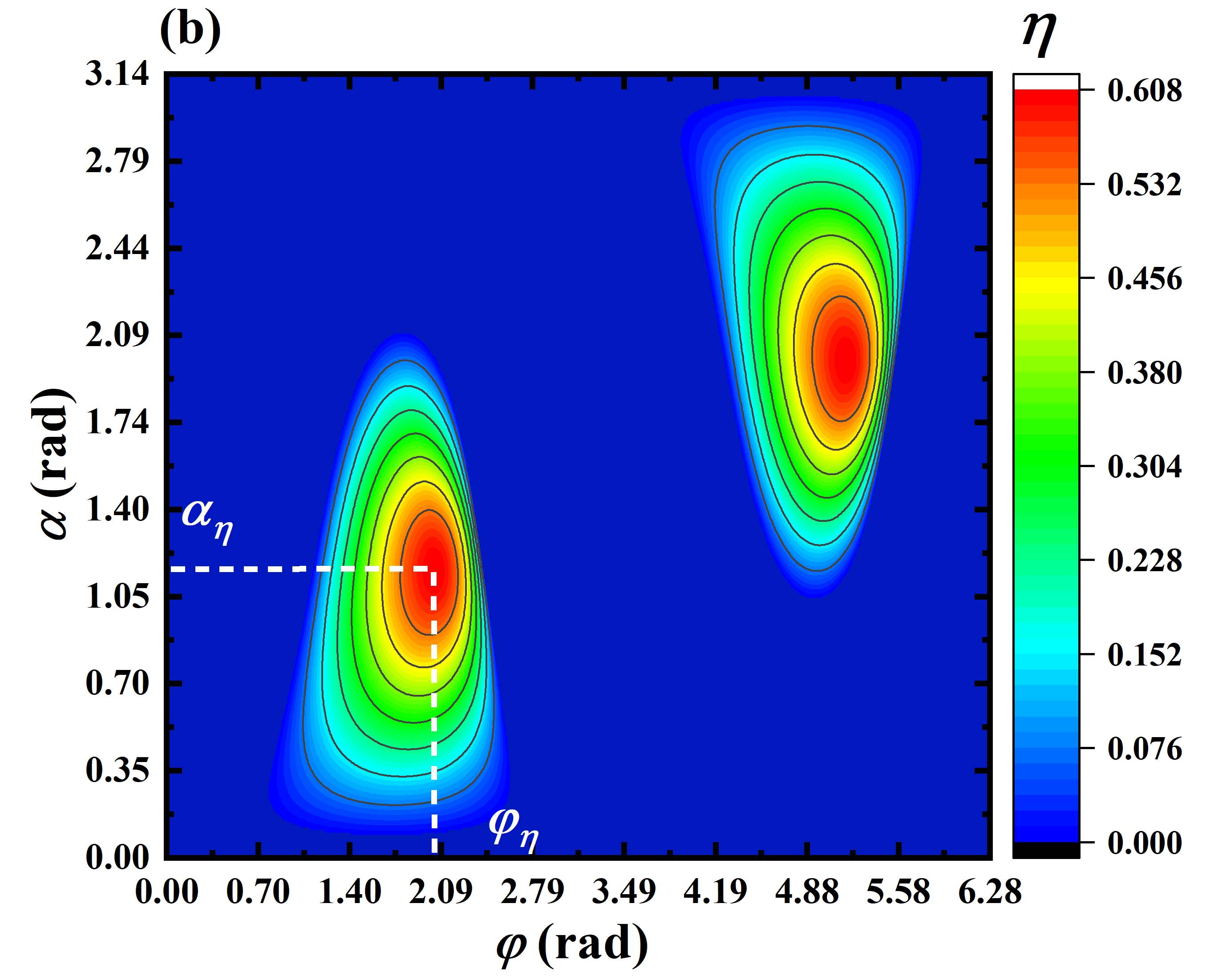}
\par\end{centering}
\caption{(a) The work output \textminus $W$ and (b) the efficiency $\eta$
varying with the colatitude $\alpha$ and longitude $\varphi$ .}
\end{figure}

A thermal divergence $D\left(\rho_{i}\|\sigma_{i}^{eq}\right)$ relates
the internal energy $\mathcal{E}_{i}$, the free energy $F_{i}^{\mathrm{eq}}$,
and von Neumann entropy $S_{i}$. The thermal divergence also measures
the difference between the nonequilibrium free energy $F_{i}$ and
equilibrium free energy $F_{i}^{\mathrm{eq}}$. To quantify the advantage
of using the concept of the thermal divergence, we first plot \textminus $W$
, $Q_{M}$, and $\eta$ {[}the thermal divergences $D\left(\rho_{3}||\sigma_{3}^{eq}\right)$
and $D\left(\rho_{4}||\sigma_{4}^{eq}\right)$, entropy difference
$\Delta S_{M}$, and nonequilibrium free energies $F_{3}$ and $F_{4}${]}
as functions of the colatitude $\alpha$, as shown in Fig. 3(a) {[}Fig.
3(b){]}. Note that the curves of $D\left(\rho_{2}||\sigma_{2}^{eq}\right)$,
$F_{1}$, and $F_{2}$ are not displayed in Fig. 3(b), because they
are constant values at a given time of the adiabatic process. 

Based on numerical simulations, we observe that \textminus $W$ increases
as $\varphi$ increases in the range of $\alpha\leq\alpha_{W^{\prime}}$.
The reason can be attributed to the enhancement of the difference
$D\left(\rho_{3}||\sigma_{3}^{eq}\right)-D\left(\rho_{4}||\sigma_{4}^{eq}\right)$
of thermal divergences. For $\alpha\geq\alpha_{W^{\prime}}$, the
difference $D\left(\rho_{3}||\sigma_{3}^{eq}\right)-D\left(\rho_{4}||\sigma_{4}^{eq}\right)$
tends to decrease, yielding low \textminus $W$ again. From the second
equality in Eq. (\ref{eq:W}) and Fig. 3(b), changing the difference
of thermal divergences is in fact modifying the discrepancy of nonequilibrium
free energies $F_{3}-F_{4}$, which is a part of the energy available
to perform thermodynamic work. The quantum heat $Q_{M}$ is closely
linked to the changes of the nonequilibrium free energy $F_{3}$ and
the entropy difference $\Delta S_{M}$ due to the measurement process
{[}Eq. (\ref{eq:QM}){]}. However, $F_{3}$ and $\Delta S_{M}$ in
the range between $\varphi_{W^{\prime}}$ and $\varphi_{\eta}$ are
relatively small, such that the cycle use fewer inputs $Q_{M}$ to
produce more output \textminus $W$. Figs. 3(c) and (d) show how the
work output and the efficiency are improved by adjusting the longitude
$\varphi$. \textminus $W$ has a maximum value at $\varphi=\varphi_{W^{\prime}}$,
which is condition for the largest values of the differences $D\left(\rho_{3}||\sigma_{3}^{eq}\right)-D\left(\rho_{4}||\sigma_{4}^{eq}\right)$
and $F_{3}-F_{4}$ as well. After reaching the maximum work output,
$D\left(\rho_{3}||\sigma_{3}^{eq}\right)$, $F_{3}$, and $\Delta S_{M}$
are reduced with the growth of $\varphi$, leading to a rapid decrease
in $Q_{M}$. Therefore, the maximum efficiency is obtained at the
point $\varphi_{\eta}$ larger than $\varphi_{W^{\prime}}$. Overall,
one can efficiently generate work from quantum engine by optimizing
the parameters with respect to the measurement basis. 

\begin{figure}
\noindent \begin{centering}
\includegraphics[scale=0.3]{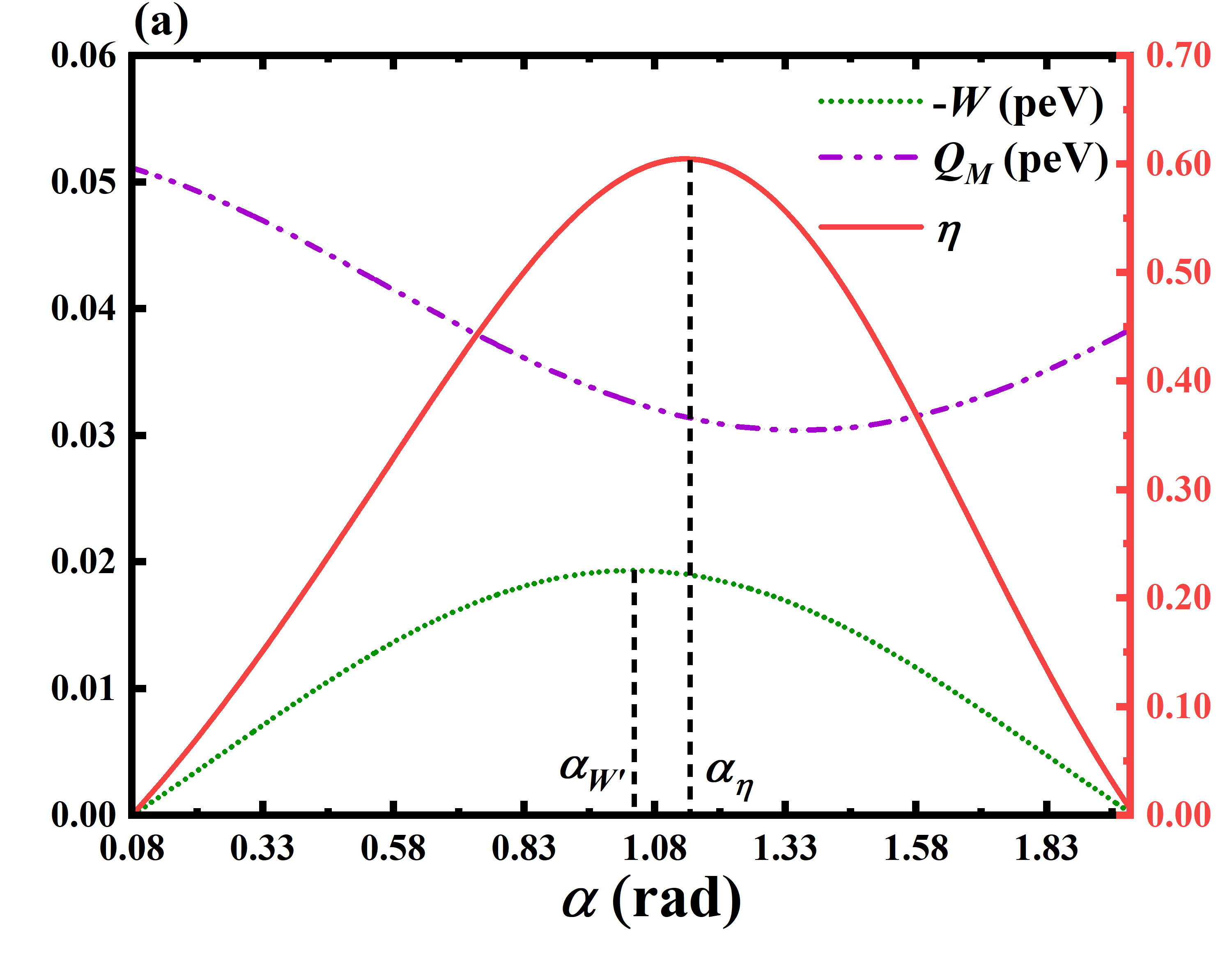}\includegraphics[scale=0.3]{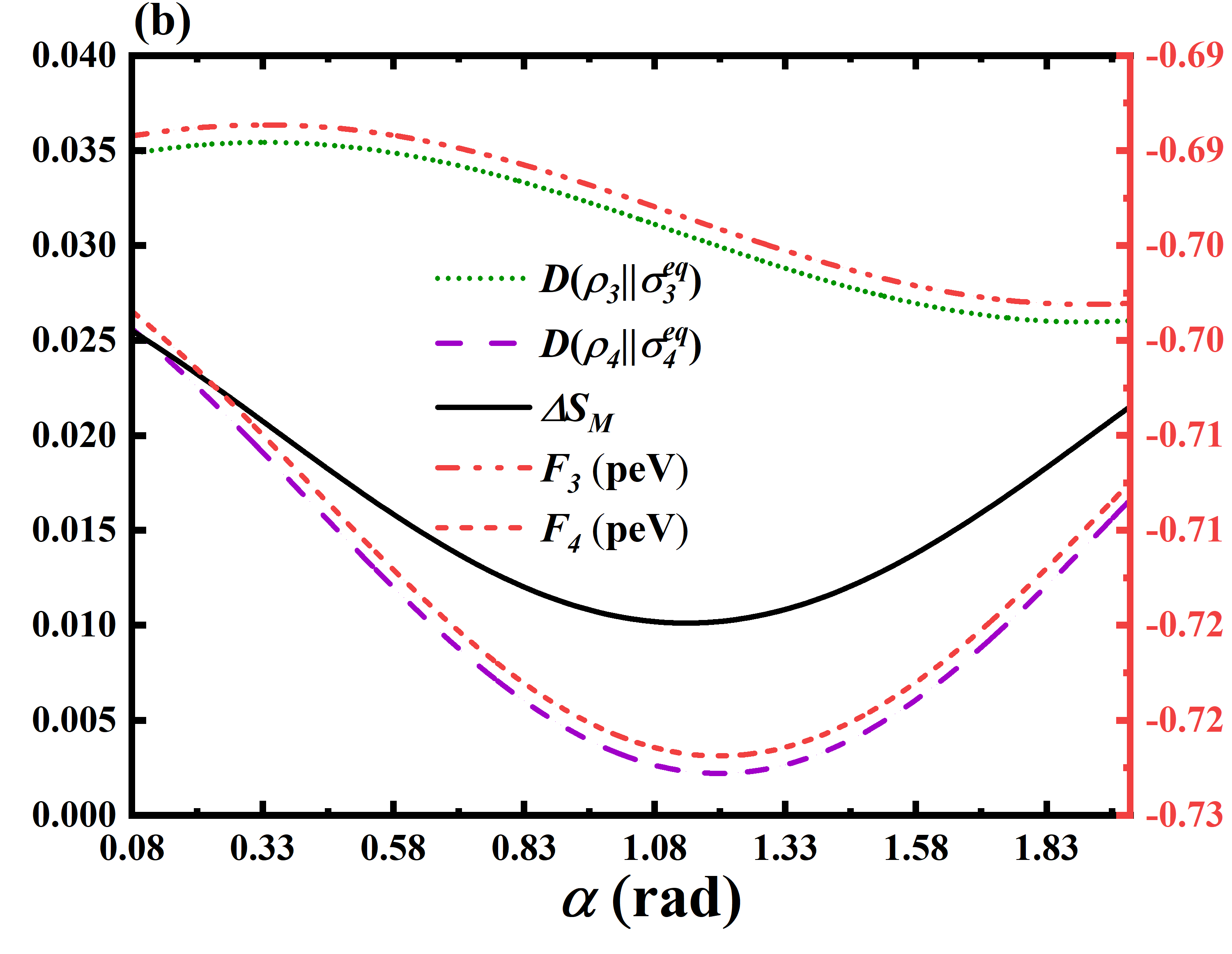}
\par\end{centering}
\noindent \begin{centering}
\includegraphics[scale=0.3]{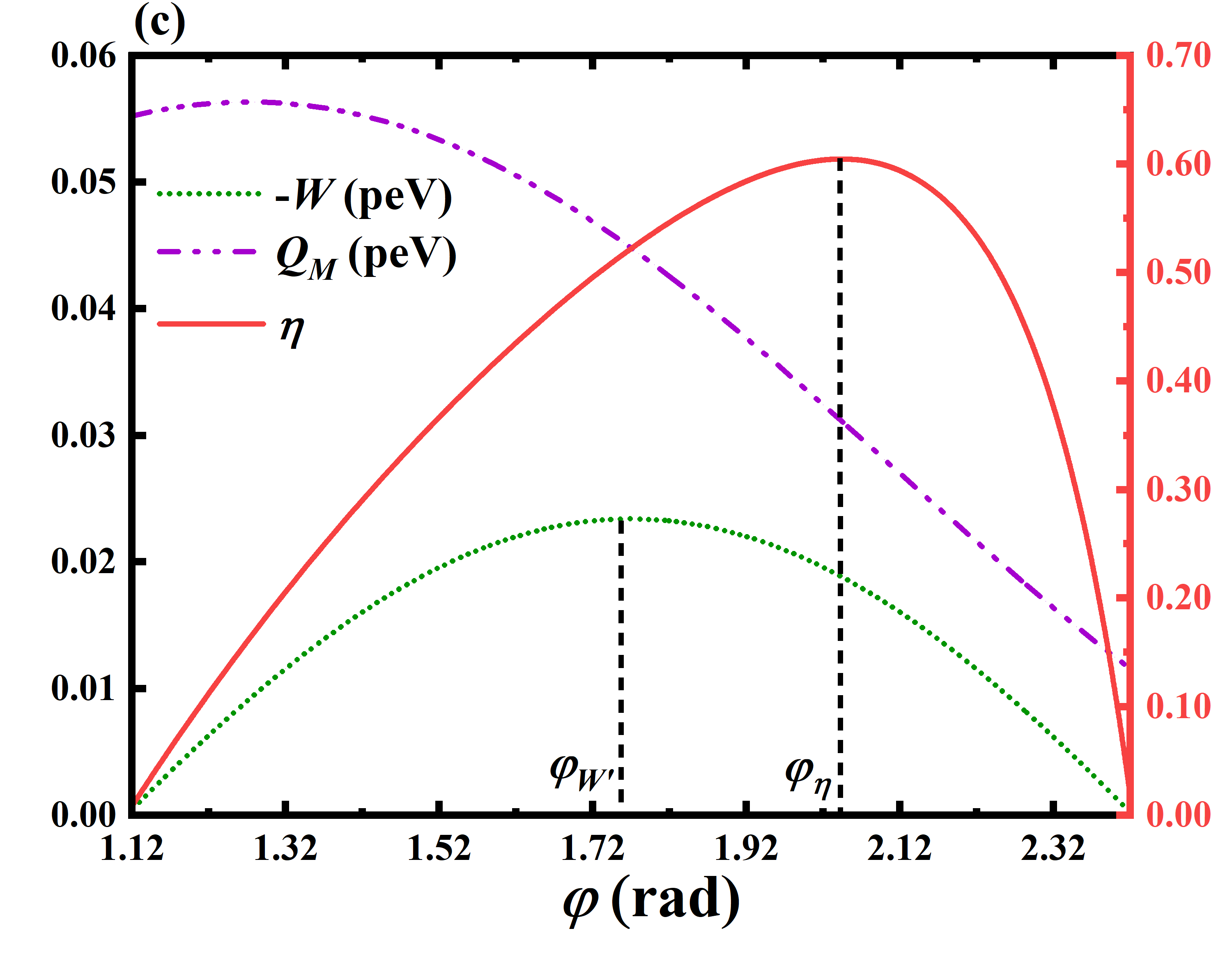}\includegraphics[scale=0.3]{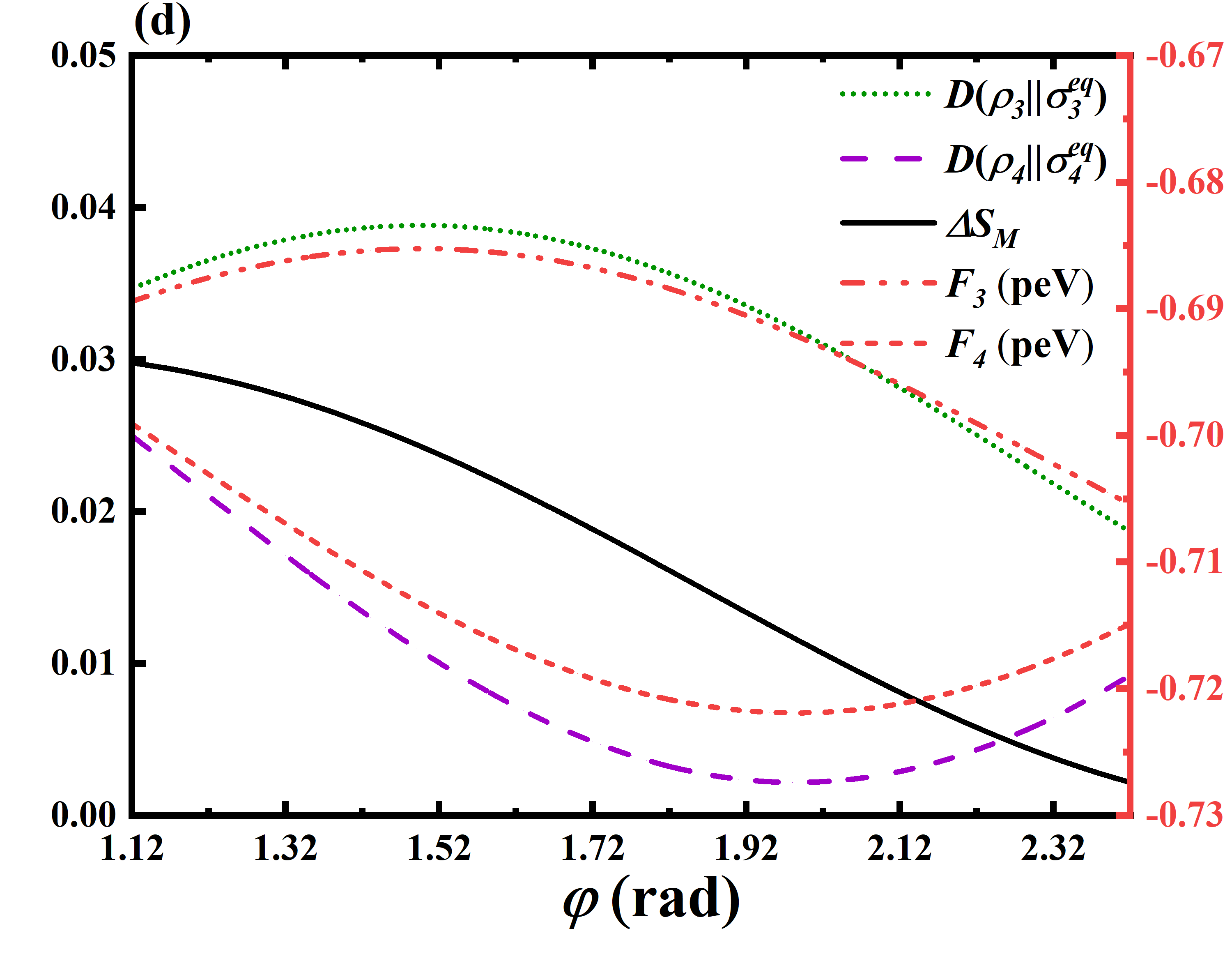}
\par\end{centering}
\caption{For a given value $\varphi_{\eta}$ of $\varphi$, the curves of (a)
the work output \textminus $W$, quantum heat $Q_{M}$, and efficiency
$\eta$ and (b) the thermal divergences $D\left(\rho_{3}||\sigma_{3}^{eq}\right)$
and $D\left(\rho_{4}||\sigma_{4}^{eq}\right)$, entropy difference
$\Delta S_{M}$, and nonequilibrium free energies $F_{3}$ and $F_{4}$
varying with the colatitude $\alpha$. For a given value $\alpha_{\eta}$
of $\alpha$, the curves of (c) the work output \textminus $W$, input
energy $Q_{M}$, and efficiency $\eta$ and (d) the thermal divergences
$D\left(\rho_{3}||\sigma_{3}^{eq}\right)$ and $D\left(\rho_{4}||\sigma_{4}^{eq}\right)$,
entropy difference $\Delta S_{M}$, and nonequilibrium free energies
$F_{3}$ and $F_{4}$ varying with longitude $\varphi$. In (a) and
(c), the left vertical axis shows values for \textminus $W$ and $Q_{M}$,
while the corresponding scale of $\eta$ is on the right vertical
axis. In (b) and (d), the left vertical axis shows values for $D\left(\rho_{3}||\sigma_{3}^{eq}\right)$,
$D\left(\rho_{4}||\sigma_{4}^{eq}\right)$, and $\Delta S_{M}$, while
the corresponding scales of $F_{3}$ and $F_{4}$ are on the right
vertical axis.}
\end{figure}

\subsection*{5. Conclusions}

In this paper, a measurement based spin engine has been developed
to show how quantum measurement is able to fuel an engine. By considering
finite-time adiabatic driven operations, the compression and expansion
of the spin transition frequency during the adiabatic stages are not
necessary for the work extraction. Analytical expressions of heat
and work are derived, highlighting the role of the thermal divergence.
We employed the relationship among the thermal divergence, nonequilibrium
free energy, and energy entropy to find how the measurement basis
can be optimized to improve the work output and the efficiency. The
results obtained here will encourage new experimental efforts to explore
engines driven by quantum measurement.
\begin{acknowledgments}
This work has been supported by the National Natural Science Foundation
of China(Grant No. 11805159 and 12075197) and the Natural Science
Foundation of Fujian Province (No. 2019J05003).
\end{acknowledgments}

\section*{APPENDIX: \textup{THE HEAT AND WORK OF EACH PROCESS}}

\subsubsection{Process I\textemdash \textemdash First adiabatic process}

The system is initially in thermal equilibrium with the thermal divergence
$D\left(\rho_{1}\|\sigma_{1}^{eq}\right)=0$, and its internal energy
is related to the free energy and the entropy of state $\rho_{1}$
as

\begin{equation}
\mathcal{E}_{1}=\frac{1}{\beta}S_{1}+F_{1}=\frac{1}{\beta}S_{1}+F_{1}^{eq}.
\end{equation}
According to Eq.(\ref{eq:div-energy-entropy}), the internal energy
of the system after the unitary evolution from $t=0$ to  $t=\tau$
is given by

\begin{equation}
\mathcal{E}_{2}=\frac{1}{\beta}\left[D\left(\rho_{2}||\sigma_{2}^{eq}\right)+S_{2}\right]+F_{2}^{eq}.
\end{equation}

\noindent The equality $S_{1}=S_{2}$ holds because of the fact that
the von Neumann entropy is invariant with respect to a unitary transformation.
Since the system is not connected to the heat bath during process
I, the work $W_{1}$ done by external field is expected to be the
change of the internal energy of the system, i.e., 

\begin{align}
W_{I} & =\mathcal{E}_{2}-\mathcal{E}_{1}\nonumber \\
 & =\frac{1}{\beta}D\left(\rho_{2}||\sigma_{2}^{eq}\right)+F_{2}^{eq}-F_{1}^{eq}.\label{eq:WI}
\end{align}

\subsubsection{Process II\textemdash \textemdash Quantum measurement}

The quantum measurement process converts the system into state $\rho_{3}$
with the internal energy 

\begin{equation}
\mathcal{E}_{3}=\frac{1}{\beta}\left[D\left(\rho_{3}||\sigma_{3}^{eq}\right)+S_{3}\right]+F_{3}^{eq}.
\end{equation}
The instantaneous projective measurement does change the Hamitonian
of the system, resulting in the free energy difference $F_{3}^{eq}-F_{2}^{eq}=0$.
Consequently, the change of the internal energy of the working substance
caused by $Q_{M}$ during quantum measurement is straightforward given
by

\begin{align}
Q_{M} & =\mathcal{E}_{3}-\mathcal{E}_{2}\nonumber \\
 & =\frac{1}{\beta}\left[D\left(\rho_{3}||\sigma_{3}^{eq}\right)-D\left(\rho_{2}||\sigma_{2}^{eq}\right)+S_{3}-S_{2}\right],
\end{align}
which can be regarded as the fuel energy for the quantum engine.

\subsubsection{Process III\textemdash \textemdash Second adiabatic process}

In the second adiabatic process, the system experiences a time-dependent
unitary evolution until reaching the internal energy

\begin{equation}
\mathcal{E}_{4}=\frac{1}{\beta}\left[D\left(\rho_{4}||\sigma_{4}^{eq}\right)+S_{4}\right]+F_{4}^{eq}.
\end{equation}
Similarly, the von Neumann entropy of the system remains constant
throughout process III, such that $S_{4}=S_{3}$. The work performed
on the system becomes

\begin{align}
W_{III} & =\mathcal{E}_{4}-\mathcal{E}_{3}\nonumber \\
 & \frac{1}{\beta}\left[D\left(\rho_{4}||\sigma_{4}^{eq}\right)-D\left(\rho_{3}||\sigma_{3}^{eq}\right)\right]+F_{4}^{eq}-F_{3}^{eq}.\label{eq:WIII}
\end{align}
With the help of Eqs. (\ref{eq:WI}) and (\ref{eq:WIII}), the total
average work $W$ done on the system is the sum of $W_{I}$ and $W_{III}$

\begin{align}
W & =W_{I}+W_{III}\nonumber \\
 & =\frac{1}{\beta}\left[D\left(\rho_{4}||\sigma_{4}^{eq}\right)+D\left(\rho_{2}||\sigma_{2}^{eq}\right)-D\left(\rho_{3}||\sigma_{3}^{eq}\right)\right],
\end{align}
where the relations $F_{3}^{eq}=F_{2}^{eq}$ and $F_{4}^{eq}=F_{1}^{eq}=F_{1}$
have been applied.

\subsubsection{Process IV\textemdash \textemdash Thermalization}

When the system is put into contact with the thermal bath, the full
thermalization process drives the system back to the initial state
$\rho_{1}$. The energy change of the system is caused by the heat
$Q_{T}$ flowing from the heat bath, i.e.,

\begin{align}
Q_{T} & =\mathcal{E}_{1}-\mathcal{E}_{4}\nonumber \\
 & =\frac{1}{\beta}\left[-D\left(\rho_{4}||\sigma_{4}^{eq}\right)+S_{1}-S_{4}\right]\text{.}
\end{align}

\end{document}